\documentclass{pasj01}

\usepackage{times}
\usepackage{amsmath,mathrsfs}
\usepackage{graphicx}
\usepackage{pdflscape}
\usepackage{color}
\usepackage{bm}

\newcommand{\bey}{\begin{eqnarray}}
\newcommand{\eey}{\end{eqnarray}}
\newcommand{\be}{\begin{equation}}
\newcommand{\ee}{\end{equation}}

\newcommand{\wt}{\widetilde}
\newcommand{\nn}{\nonumber}

\begin{document}

\title{The Subaru FMOS galaxy redshift survey (FastSound). V. Intrinsic alignments of emission line galaxies at $z\sim 1.4$}
\author{
Motonari \textsc{Tonegawa},\altaffilmark{1,2,*}
Teppei \textsc{Okumura},\altaffilmark{3,4,*} 
Tomonori \textsc{Totani},\altaffilmark{5,6}  \\
Gavin \textsc{Dalton},\altaffilmark{7,8}
Karl \textsc{Glazebrook},\altaffilmark{9}
and Kiyoto \textsc{Yabe}\altaffilmark{4}}

\altaffiltext{1}{Korea Institute for Advanced Study, 207-43 Cheongnyangri-dong, Dongdaemun-gu, Seoul, Korea}
\altaffiltext{2}{Institute of Astronomy, National Tsing Hua University, No. 101, Section 2, Kuang-Fu Road, Hsinchu, Taiwan 30013}
\altaffiltext{3}{Institute of Astronomy and Astrophysics, Academia Sinica, P. O. Box 23-141, Taipei 10617, Taiwan}
\altaffiltext{4}{Kavli Institute for the Physics and Mathematics of the Universe, The University of Tokyo,
5-1-5 Kashiwanoha, Kashiwa, Chiba 277-8583, Japan}
\altaffiltext{5}{Department of Astronomy, School of Science, The University of Tokyo, 7-3-1 Hongo, Bunkyo-ku, Tokyo 113-0033, Japan}
\altaffiltext{6}{Research Center for the Early Universe, School of Science, The University of Tokyo, 7-3-1 Hongo, Bunkyo-ku, Tokyo 113-0033}
\altaffiltext{7}{Department of Astrophysics, University of Oxford, Keble Road, Oxford OX1 3RH, UK}
\altaffiltext{8}{STFC RALSpace, Harwell, Oxfordshire OX11 0QX, UK}
\altaffiltext{9}{Centre for Astrophysics \& Supercomputing, Swinburne University of Technology, P.O. Box 218, Hawthorn, VIC 3122, Australia}

\email{tonegawa@kias.re.kr, tokumura@asiaa.sinica.edu.tw}


\maketitle
 
\begin{abstract}
Intrinsic alignments (IA), the coherent alignment of intrinsic galaxy orientations,
can be a source of a systematic error of weak lensing surveys.
The redshift evolution of IA also contains information about the physics of galaxy formation and evolution.
This paper presents the first measurement of IA at high redshift, $z\sim 1.4$,
using the spectroscopic catalog of blue star-forming galaxies of the FastSound redshift survey, 
with the galaxy shape information from the Canada-Hawaii-France telescope lensing survey.
The IA signal is consistent with zero with power-law amplitudes fitted to the projected correlation functions
for density-shape and shape-shape correlation components,
$A_{\delta+}=-0.0071\pm 0.1340$ and $A_{++}=-0.0505\pm 0.0848$, respectively.
These results are consistent with those obtained from blue galaxies at lower redshifts
(e.g., $A_{\delta+}=0.0035_{-0.0389}^{+0.0387}$ and $A_{++}=0.0045_{-0.0168}^{+0.0166}$ at $z=0.51$ from the WiggleZ survey).
The upper limit of the constrained IA amplitude corresponds to a few percent contamination to the weak-lensing shear power spectrum,
resulting in systematic uncertainties on the cosmological parameter estimations by 
$-0.052<\Delta \sigma_8<0.039$ and $-0.039<\Delta \Omega_m<0.030$.
\end{abstract}

\section{Introduction}\label{sec:intro}
Galaxies are one of the most fundamental objects which can be treated as a tracer of the large-scale structure of the Universe.
Moreover, shapes of galaxies contain the information about how the galaxies formed and evolved (e.g., \cite{Binney:1981,Lambas:1992}). 
Galaxies are formed through the gravitational accumulation of gas induced by the density fluctuation of dark matter
(\cite{Binney:1977}; \cite{Tegmark:1997}; \cite{Benson:2010}),
then they experience internal and external effects
which can produce the coherent alignment of galaxy orientations, the so-called intrinsic alignments (IA).
For example, tidal forces exerted by the gravitational field of dark matter halos or a large-scale structure
align galaxies along a particular direction (\cite{Ciotti:1994}; \cite{Pereira:2005}; \cite{Kuhlen:2007}).
On the other hand, halo/galaxy mergers can randomize the shape of galaxies, erasing galaxy alignments.
Feedback processes caused by supernovae and AGNs may change the spatial distribution of the stellar components inside galaxies (\cite{Okamoto:2005}; \cite{Scannapieco:2008}; \cite{Dubois:2016}),
modifying the optically-observed galaxy shapes on images.
The detail study of IA would enable us to test models of galaxy formation and evolution.

IA also play an important role in a cosmological context. 
In weak gravitational lensing, 
light rays are bent by the gravitational field of the large-scale structure, resulting in distorted images of galaxies, known as the gravitational shear 
\citep{Bartelmann:2001}.
Since the gravitational shear provides a direct measurement of the matter distribution,
it enables us to constrain cosmological parameters such as the matter density parameter $\Omega_{\rm m}$,
the amplitude of density fluctuation $\sigma_8$, and the equation of state of dark energy $w_0$.
\citet{Kilbinger:2013} analyzed the data from Canada-France-Hawaii Telescope lensing survey (CFHTLenS: \cite{Heymans:2012}) and constrained cosmological parameters.
Ongoing and future lensing surveys, such as the Dark Energy Survey (DES: \cite{The-Dark-Energy-Survey-Collaboration:2005}), Hyper Suprime-Cam (HSC: \cite{Miyazaki:2012}) and Large Synoptic Survey Telescope
(LSST: \cite{LSST-Dark-Energy-Science-Collaboration:2012})
will provide even tighter cosmological constraints. 
However, the observed shear is always the sum of the lensing signal $\gamma^{G}$ and the intrinsic shape $\gamma^{I}$, namely $\gamma = \gamma^G + \gamma^I$.
If the intrinsic shapes of galaxies had random orientations, the power spectrum of the observed ellipticities would be equal to the gravitational shear power spectrum, $\langle \gamma\gamma\rangle = \langle \gamma^{G}\gamma^{G}\rangle$.
Under the presence of IA, however, the contamination of the $\gamma^I$ term does not disappear and we need to carefully model the contamination; otherwise we would obtain the biased cosmological constraints.

There are two types of IA contaminations to the gravitational shear analysis:  
the ellipticity correlation of neighboring galaxies (intrinsic ellipticity-ellipticity correlation) and the correlation of intrinsic galaxy ellipticities with the density field responsible for lensing shear (gravitational shear-intrinsic ellipticity correlation), respectively known as the II and GI correlations (see \cite{Schafer:2009,Joachimi:2015,Kirk:2015} for the reviews).
Theoretical modeling of the II correlation has been done by a lot of work using both analytical and numerical methods 
(e.g., \cite{Heavens:2000,Croft:2000,Catelan:2001,Jing:2002,King:2002,Takada:2004}). 
The II term is generated by physically close galaxy pairs which are subject to the tidal field of the same dark matter structure and thus it is relatively straightforward to model and subtract the II contamination.
On the other hand, the GI term is harder to subtract because the GI correlation exists for the galaxy pairs physically distant along the line of sight (\cite{Hirata:2004}, hereafter HS04). 
Thus precisely modeling the GI contamination is one of the most important tasks to use weak lensing for precision cosmology. 
There are several attempts to model the GI effect, based on the self-calibration technique using the II correlation \citep{Okumura:2009a,Zhang:2010}, (non)linear alignment models \citep{Bridle:2007,Joachimi:2011,Blazek:2011} and a halo model \citep{Schneider:2010}. 
The IA contaminations have been also studied using large hydrodynamic simulations, taking into account the astrophysical effects (e.g., \cite{Tenneti:2015,Chisari:2015}).

Both the II and GI correlations have been measured in various galaxy surveys (e.g., \cite{Pen:2000,Heymans:2004,Mandelbaum:2006}; HS04; \cite{Okumura:2009,Faltenbacher:2009,Joachimi:2011,Mandelbaum:2011,Singh:2015,van-Uitert:2017}), and the clear signals have been detected for early-type galaxies (HS04; \cite{Okumura:2009,Singh:2015}), which are in good agreement with the linear alignment model as well as the $\Lambda$CDM model with a Gaussian misalignment between the major axes of galaxies and their host halos.
On the other hand, IA of late-type galaxies have not been detected yet, even in the latest observation \citep{Mandelbaum:2011}.
It may be due to the large angular momentum of spiral galaxies or galaxy mergers, which may have disrupted IA which have existed at high redshifts. On the other hand, it is also possible that IA are not present for late-type galaxies at all the epochs.
Therefore, measuring IA for late-type galaxies at higher redshifts is important to fully understand the galaxy formation process. 
Moreover, the effect of IA at higher redshifts will be crucial for deeper lensing surveys because the fraction of late-type galaxies will be larger in the earlier Universe.
Nevertheless, so far the observational studies of IA have been limited to $z<1$. 

In this paper, we present the first measurement of IA at $z>1$ using the galaxies obtained from the FastSound galaxy survey. We then put a constraint on the contamination of IA to
the cosmological parameters estimation in weak lensing surveys.
FastSound is a redshift survey for emission line galaxies with H$\alpha$
and has collected $\sim4000$ redshifts at $1.2<z<1.6$ (Paper I: \cite{Tonegawa:2015}) using the Subaru Telescope. 
The shape information of the FastSound galaxies is taken from the CFHTLenS data, which provides accurate shape measurements dedicated to lensing studies.

The structure of this paper is as follows.
In section \ref{sec:theory} we briefly describe the theoretical formalism of weak lensing and IA. 
In section \ref{sec:data}, we describe the FastSound and CFHTLenS data used in this work.
Section \ref{sec:measurement} presents the measurements of the GI and II correlation functions,
followed by the constraints on IA and its cosmological implications on the shear measurement of weak lensing survey in section \ref{sec:constraints}.
We conclude in section \ref{sec:conclusion}.
Unless otherwise stated, we adopt a standard set of the cosmological parameters:
$(\Omega_m,\Omega_\Lambda,h,\sigma_8) = (0.3,0.7,0.7,0.8)$, where $h$ is the normalized Hubble constant $H_0/(100\,{\rm km/s/Mpc})$.

\section{Formalism}\label{sec:theory}

In this section, we briefly summarize the quantities in weak lensing surveys which are related to the statistics of IA, following the formulation presented in HS04 and \citet{Mandelbaum:2006}. 
We assume a flat universe in this paper, but the generalization of the formalism to the curved universe
is straightforward and can be found in HS04 and \citet{Mandelbaum:2006}.

Since $\gamma = \gamma^G + \gamma^I$ as mentioned in section \ref{sec:intro}, 
the cross-power spectrum of the observed shear between $i$-th and $j$-th redshift bins
can be decomposed into three components (HS04):
\begin{equation}
C^{(ij)}(\ell) = C_{\rm GG}^{(ij)}(\ell) + C_{\rm GI}^{(ij)}(\ell) + C_{\rm II}^{(ij)}(\ell),
\label{equation:Cij}
\end{equation}
where $\ell=k\chi$, $k$ is the wavenumber and $\chi(z)$ is the comoving distance at redshift $z$. 
The first term is the gravitational shear power spectrum from which cosmological information is extracted, 
and the second and third terms are respectively the GI and II correlations, which are our main interests and can be a source of systematics on weak lensing cosmology. 
Each term is related to the underlying power spectrum as
\bey
&&C_{\rm GG}^{(ij)}(\ell)
=\!\! \int_0^\infty \!{
q_i(\chi) q_j(\chi)
\over\chi^2}
P_\delta(k;\chi)
d\chi, \label{eq:c_gg}\\
&&C_{\rm II}^{(ij)}(\ell)
=\!\! \int_0^\infty \!{
n_i(\chi) n_j(\chi)
\over\chi^2}
P_{\tilde \gamma^{\rm I}} (k;\chi)
d\chi,\label{eq:c_ii}\\
&&C_{\rm GI}^{(ij)}(\ell)
=\!\! \int_0^\infty \!{
q_i(\chi) n_j(\chi)
\over\chi^2}
P_{\delta, \tilde \gamma^{\rm I}}(k;\chi)
d\chi, \label{eq:c_gi}
\eey
where $P_\delta(k;\chi)$ is the power spectrum of the matter density fluctuation $\delta$,
$P_{\tilde \gamma^{\rm I}}(k;\chi)$ is that of the galaxy density-weighted intrinsic shear
$\tilde \gamma^{\rm I} =(1+\delta_g)\gamma^{\rm I}$ with the galaxy density fluctuation $\delta_g$,
and $P_{\delta, \tilde \gamma^{\rm I}}(k;\chi)$ is the cross-power spectrum between $\delta$ and $\tilde \gamma^{\rm I}$.

$n_i(\chi)$ is the normalized galaxy distribution of $i$-th bin, and 
\begin{equation}
q_i(\chi) =
{3\over 2}\Omega_m
\frac{H_0^2}{c^2}
(1+z)
\int_0^\infty
\!\!\!\!\!
n_i(\chi')
\frac{(\chi'-\chi) \chi}{\chi'}
d\chi'.
\label{eq:wint}
\end{equation}
In this paper, we only consider the linear alignment model (\cite{Catelan:2001}; HS04),
which relates the matter power spectrum to these two power spectra linearly (see section \ref{sec:constraints}).

These two IA power spectra can be related to the projected correlation functions, which we want to measure from the observations, as \citep{Bridle:2007}:
\bey
&& w_{\delta+}(r_p) = -\frac{1}{2\pi} \int
P_{\delta, \tilde \gamma^{\rm I}}
 J_{2}(kr_p)kdk,
\label{equation:wd+} \\
&& w_{++}(r_p) = \frac{1}{2\pi} \int
P_{\tilde \gamma^{\rm I}}
 J_{0}(kr_p)kdk,
\label{equation:w++}
\eey
where $J_{\,n}(kr_p)$ is the Bessel function of the first kind of $n$-th order,
$r_p$ is the transverse separation, and
$w_{\delta+}(r_p)$ represents the correlation between the intrinsic ellipticity of a galaxy and matter overdensity. 
If the linear galaxy bias, $b_g$, is assumed, the GI correlation of galaxies
$w_{g+}(r_p)$ is simply related to $w_{\delta+}(r_p)$ by 
\begin{equation}
w_{g+}(r_p)=b_g w_{\delta+}(r_p).\label{equation:lbias}
\end{equation}

In the following, we describe the estimator and measurement of $w_{g+}$ and $w_{++}$,
which are connected to the GI and II power spectrum.
The estimate of $C_{\rm GI}$ and cosmological implications are given in section \ref{sec:constraints}.


\section{Data}\label{sec:data}
\subsection{FastSound spectroscopic sample}
FastSound is a near-infrared spectroscopic survey of star-forming
galaxies at $z = $ 1.19--1.55.  It used the fiber-multi object
spectrograph (FMOS: \cite{Kimura:2010}) mounted on the Subaru Telescope,
which was able to obtain $\sim400$ spectra within a radius of $15$
arcmins simultaneously.  The primary scientific goal of FastSound was
to measure the structure growth rate at such a high redshift by using
the redshift space distortion (RSD) effect.  This gives a test of
general relativity as a theory of gravity on cosmological scales,
which is important to understand the origin of the accelerated
expansion of the universe. The main results about RSD cosmological
implications were already published (Paper IV: \cite{Okumura:2016}).

The target galaxies are selected using the photometric redshifts and
H$\alpha$ fluxes estimated by spectral energy distribution
(SED) fitting applied to five optical magnitudes of the
Canada-France-Hawaii Telescope legacy survey (\cite{Gwyn:2012}), to
select bright H$\alpha$ emission-line galaxies at such high redshifts.
Observations were carried out from April 2012 to July 2014, covering
$\sim25$ deg$^2$ over the CFHTLS Wide W1--4 fields in total.  Due to the
variation of weather conditions, there is a substantial difference in
the observed regions; 10, 39, 54, and 18 field-of-views (FoVs) in the
W1--4 fields, respectively.  The FMOS images were processed by the
standard pipeline (\textit{FIBER-pac}: \cite{Iwamuro:2012}) to produce
2-D reduced images, which were then passed to an automated
emission-line detection algorithm (\textit{FIELD}:
\cite{Tonegawa:2015a}) to create the redshift catalog
(Paper I), which is used in this work.  In this study, we
use the FastSound spectroscopic data in CFHTLS W2 and W3 sub-fields.
Each sub-field is centered at $(\alpha, \delta)=(134.5, -3.2)$ and
$(214.5, 53.2)$ and covers $8$ deg$^2$ and $11$ deg$^2$,
respectively. Following Paper IV, some FMOS FOVs were removed from the analysis
when either of the two spectrographs (IRS1 and IRS2) was not working,
or the observing condition was poor. 

The line signal-to-noise ($S/N$) threshold of $4.5$ is applied to the
spectroscopic catalog, yielding $1,265$ and $1,175$ galaxies for W2
and W3 respectively.  The false detection rate $f_{\rm fake}$ is
estimated to be $f_{\rm fake}=0.041$ (\cite{Tonegawa:2015a};
Paper IV) at this S/N threshold. 
Because of the relatively narrow wavelength coverage
($1.44$--$1.68\mu$m) of FastSound, only one emission line is detected
in spectra of most galaxies. Since H$\alpha$ is the predominantly
bright emission line in this wavelength range, redshift is calculated
assuming that these are H$\alpha$. Spectroscopic line identification
was possible for a small fraction of bright galaxies by multiple
emission lines detected in their spectra (see Paper II: \cite{Okada:2016}).
Reliability of the line identification in the sample was studied in detail
in Paper II.  Misidentifications are predominantly
caused by identifying the stronger line of the [OIII] doublet as
H$\alpha$, and its probability is estimated to be $f_{\rm
  [OIII]}=0.032$.  Contamination due to the detections of false lines and
non-H$\alpha$ lines leads to the total redshift blunder rate of
$f_{\rm blund}=0.071$, which causes a change of the correlation
function amplitudes.  We take into account this effect by multiplying
the model predictions by $(1-f_{\rm blund})^2$ in section \ref{sec:constraints}.

Physical properties of the FastSound galaxy sample were studied in
detail by Paper II and III (\cite{Yabe:2015}). 
Typical ranges of the stellar mass and star formation rate are $10^9
\rm{M_\odot}$--$10^{11} \rm{M_\odot}$ and $50$--$1000
\rm{M_\odot/yr}$, respectively, as estimated by SED fitting to the
photometric data of CFHTLS (optical),
which are combined with UKIDSS (near-infrared, \cite{Lawrence:2007}) and
Spitzer/IRAC (mid-infrared) data if these are available.

A random catalog, which has the same sky and redshift coverage as the spectroscopic data, is needed to calculate the correlation functions.
We use the random catalog created by Paper IV, which was used for the RSD analysis of FastSound galaxies.
Random points are distributed to match the angular selection function and the radial selection function.
We construct the random catalog with the number density 20 times as large as the data catalog.

\subsection{Shape measurement}\label{sec:shape}
We need shape information for the FastSound galaxy sample to study IA, and we use the CFHTLenS data (\cite{Heymans:2012}) for this.
The CFHTLenS covers $154$ deg$^2$ of the sky in five optical bands $u^*$,$g'$,$r'$,$i'$,and $z'$ with the MegaCam camera.
The procedure to create the catalog is detailed in \citet{Erben:2013},
and all the shape data used in our study are taken from the CFHTLenS website.\footnote{http://www.cadc-ccda.hia-iha.nrc-cnrc.gc.ca/en/community/CFHTLens/query.html}
A software code, SExtractor, is run in the stacked $i'$-band images to obtain celestial coordinates $(\alpha,\delta)$ and position angle $\theta$ of objects.
Then stars are selected manually for each field-of-view to estimate the spatially varying point spread function (PSF).
To obtain ellipticities for galaxies, $(e_1, e_2)$,
the two-component model (bulge and disk) is convolved with the PSF and fitted to the objects brighter than $24.7$ mag in $i'$-band images,
followed by the Bayesian marginalization over nuisance parameters of galaxy position, size, brightness and bulge fraction,
eliminating uncertainties of these parameters for faint galaxies (\cite{Miller:2013}).
In principle, $\theta$ and $(e_1, e_2)$ should be related as:
\begin{equation} \label{eq:e1e2}
\left(
    \begin{array}{c}
      e_1  \\
      e_2 
    \end{array}
  \right)=  
  \frac{a-b}{a+b}
  \left(
    \begin{array}{c}
      \cos{2\theta}  \\
      \sin{2\theta} 
    \end{array}
  \right),
\end{equation}
where $a$ and $b$ are the semi major and minor axes respectively,
but there is a small deviation from equation (\ref{eq:e1e2})
for our data.
This is not due to the shape measurement noise, but due to the different algorithms and assumed shapes to measure $\theta$ and $(e_1, e_2)$ at different processing stages:
SExtractor (\cite{Bertin:1996}) is used for $\theta$ assuming an elliptical shape, while \textit{lensfit} (\cite{Miller:2007}) is used for $(e_1, e_2)$
assuming the two-component model.

In this study, we use $(e_1, e_2)$ for main results,
because the assumed galaxy models are realistic and
biases are well examined through various tests by the CFHTLenS team. 
However, we will also perform the analysis by combining $\theta$ and $(e_1, e_2)$ as a systematic test (see equation (\ref{equation:eplus_theta}) below).

We perform a cross-match between FastSound galaxies and CFHTLenS galaxies with a matching radius of $1.0''$,
which yields $523$ and $635$ matched objects in the W2 and W3 fields respectively,
giving the number density of $\sim90/{\rm deg^2}$.
Thus, 1158 out of the 2440 FastSound galaxies have the shape information. 
The relatively small fraction of matched objects is likely
to be due to the threshold ($S/N_{i'}>10$) applied to CFHTLenS objects for secure determination of shape parameters (\cite{Miller:2013}).

We correct for the multiplicative and additive biases in the ellipticity $(e_1,e_2)$
based on the prescription of \citet{Heymans:2012}. 
The additive bias correction is applied to the $e_2$ component on a galaxy-by-galaxy basis
using the $c_2$ value provided in the CFHTLenS catalog.
For the multiplicative bias, on the other hand, \citet{Miller:2013} argue that this approach is inaccurate
and the correction must be performed as an ensemble average.
Therefore we obtain the average of $S/N$ and size in the $i'$-band image for the $1158$ galaxies from the CFHTLenS catalog,
to calculate the correction factor by equation (14) of \citet{Miller:2013}.
The correction factor is $0.97$ and higher than the reported value (\cite{Miller:2013}) for the full CFHTLenS sample,
which is because the galaxies of our sample have higher $S/N$ in the $i'$-band images than general galaxies in the CFHTLenS catalog on average.
The formula of \citet{Miller:2013} is derived from their simulation,
where they distributed galaxies in the images using the two-component model with the distributions of physical parameters being matched to the local galaxy population.
Therefore there could be a bias on our multiplicative correction factor because the simulated galaxies do not adequately represent the shape of FastSound galaxies.
However, the convolution with the PSF makes the model choice less important especially for the high-redshift galaxies
because they are small and marginally resolved by the CFHTLenS imaging.
The issue will be more important when we use larger spectroscopic samples or images from space telescopes for better precision.

We need a random catalog for the galaxy distribution with the shape information.
We calculate the ratio between the redshift distributions of the FastSound galaxies with and without shape information for each of W2 and W3.
Using this ratio as a weight, we obtain the random catalog for the galaxy sample with shapes by drawing from the random points described in the previous section. 

There is a possibility that $f_{\rm fake}$ of the shape catalog might be different from that of the density catalog,
because the fake emission-line objects may drop during the cross-match with the CFHTLenS catalog.
To check this, we have also performed a cross-matching between the FastSound inverted catalog and the CFHTLenS data.
The inverted catalog is obtained by applying \textit{FIELD} to the inverted frames, which are created from ``sky - object'' images
rather than ``object - sky'' images in the reduction process of \textit{FIBRE-pac}.
All objects detected by \textit{FIELD} in the inverted catalog should be fake,
because real emission lines become negative in the inverted frames.
Also, the number of objects in the inverted catalog is expected to be the same as that of fake objects contaminating the density catalog (\cite{Tonegawa:2015a}).
We find that the matching probability is constant ($40$--$60\%$)
in a wide range of line $S/N$ ($3.0$--$7.0$) for both the normal and inverted catalogs,
which means that the fake objects are not eliminated by the cross-matching with CFHTLenS data,
and hence $f_{\rm fake}$ will be the same for the shape and density catalogs.
Therefore we will use a single value of $f_{\rm fake}$ to correct for redshift blunders in both the GI and II correlations.

\section{Measurement}\label{sec:measurement}
In this section, 
we first present the estimators for the GI and II correlations in section \ref{subsection:estimator}.
Its covariance matrix is presented in section \ref{sec:covariance}. Our measurements are presented in section \ref{sec:results} and their systematic effects are discussed in section \ref{subsection:systematics}. Some relevant, complimentary statistics for the GI correlation are described in section \ref{subsection:alignment_statistics}.

\subsection{Estimators}\label{subsection:estimator}
The estimator for the GI correlation was presented in \citet{Mandelbaum:2006}, by extending the Landy-Szalay estimator \citep{Landy:1993} for the density correlation function, as
\begin{equation}\label{equation:GI}
\xi_{g+}({\bf r}) = \frac{S_+(D-R)}{R_sR}, 
\end{equation}
where ${\bf r}$ is the separation vector between two points 
and $S_+D({\bf r})$ means the sum of the + component of the shear of 
$j$-th galaxy with a shape measurement relative to $i$-th galaxy at separation ${\bf r}$,
\begin{equation}\label{equation:sd}
S_+D({\bf r}) = \sum_{i\neq j| {\bf r}} w_j e_+(j|i),
\end{equation}
where  $e_+(j|i)$ is the ellipticity defined relative to the direction to the $i$-th galaxy, 
\begin{equation}\label{equation:eplus}
e_+(j|i)=-e_1\cos(2\phi)-e_2\sin(2\phi),
\end{equation}
where $\phi$ is the angle along the line joining the two galaxies measured relative to the axis the same as that used to determine $(e_1,e_2)$. 
When we also use the information of the position angle $\theta$ to determine $e_+(i|j)$, we use the expression of 
\be
e_+(j|i)=-e\cos{2(\theta-\phi)}, \label{equation:eplus_theta}
\ee
where $e=\sqrt{e_1^2+e_2^2}$. 
The uncertainty of the shape measurement for $j$-th galaxy is taken into account by the normalized weight factor $w_j$ in equation \eqref{equation:sd}, which is obtained from the CFHTLenS catalog.
$S_+R$ is measured in the same manner but $i$-th point is drawn from the random catalog with the same survey geometry as the galaxy catalog.
$R_sR$ is the pair number of random catalogs, where $R_s$ and $R$ are drawn from the random catalog
corresponding to $S_+$ and $D$ respectively.
The terms $S_+R$ and $R_sR$ are rescaled to match $S_+D$.

Equation \eqref{equation:GI} gives the preference of the galaxy orientation
toward the overdense regions; if $\xi_{g+}<0$ and $\xi_{g+}>0$, in our definition the major axes of galaxy shapes respectively tend to be aligned parallel and perpendicular to the line connecting to another galaxy, while 
$\xi_{g+}=0$ corresponds to the case where galaxy shapes have no preferred orientation. 
As explained in section \ref{sec:shape}, we use the galaxies whose shape information is provided by the CFHTLenS data as our shape catalog.
On the other hand, we use the whole FastSound sample as a representative of the density field $D$ for the main analysis.
In section \ref{subsection:systematics} we will also perform the analysis where we use the same data as the shape sample as the density field for a systematic test,
in which case $R_s$ and $R$ in equation (\ref{equation:GI}) become equivalent. 

We will also measure $\xi_{g\times}$ by replacing equation (\ref{equation:eplus}) by 
\be
e_\times (j|i) = e_1 \sin{2\phi}-e_2 \cos{2\phi}, 
\ee
estimating $S_\times $, and replacing $S_+$ by it in equation (\ref{equation:GI}). 
Since $\xi_{g\times}$ should be zero due to party symmetry, we can use it for testing systematics such as shape measurement errors. 

The II correlation, the auto-correlation of the intrinsic shapes, can be measured by the estimator:
\begin{equation}
\xi_{++}({\bf r}) = \frac{S_+S_+}{R_s R_s} \ \ \ {\rm and} \ \ \ \xi_{\times\times}({\bf r}) = \frac{S_\times S_\times}{R_s R_s},
\end{equation}
where
\begin{equation}
S_{+}S_{+}({\bf r}) =  \sum_{i\neq j| {\bf r}} w_i w_j e_+(j|i)e_+(i|j)
\end{equation}
and $S_\times S_\times$ can be computed likewise. 
We use only the galaxies whose shapes are measured from CFHTLenS data.
Just like $\xi_{g\times}$, we can measure the cross correlation, $\xi_{+\times}$, and use it for a systematic test since the quantity should be zero at all scales. 

The projected GI and II correlation functions are obtained by adopting the separation bin ${\bf r}=(r_p,r_\pi)$, where 
$r_p$ and $r_\pi$ are the separations perpendicular and parallel to the line of sight, and integrating $\xi_{AB}(r_p,r_\pi)$ along the line-of-sight;
\begin{equation}\label{equation:GI_proj}
w_{AB}(r_p)=\int \xi_{AB}(r_p,r_\pi)dr_\pi, 
\end{equation}
where $AB=\{g+,\ g\times,\ ++,\ \times\times,\ +\times \}$. 
The galaxy position-intrinsic shape correlation is known to be affected by the peculiar velocity of galaxies, i.e., RSD \citep{Kaiser:1987}, even in linear theory \citep{Singh:2015}, while the II correlation is not. 
By taking the projection, however, the effect of RSD is suppressed to be much smaller than the current statistical uncertainties.
This integration is performed from $r_\pi=-60$ to $r_\pi=60$ $h^{-1}$ Mpc,
and by changing the range we confirmed that the final results do not change very much with this choice.

\subsection{Covariance matrix}
\label{sec:covariance}

We adopt a jackknife resampling method to estimate the covariance matrix. 
We separate the FastSound W2 and W3 fields into $N$ sub-regions on the sky, and
calculate the GI correlations $N$ times, omitting each sub-region.
Then the covariance matrix for the statistics $w_{AB}$, where $AB=\{g+,\ g\times,\ ++,\ \times\times,\ +\times \}$, is obtained as 
\bey
C_{ij} = \frac{N-1}{N}
\sum_{k=1}^{N} &&[w_{AB}^k(r_{p,i})-\overline{w}_{AB}(r_{p,i})] \nn \\ 
&&\times 
[w_{AB}^k (r_{p,j})-\overline{w}_{AB} (r_{p,i})], \label{eq:covariance}
\eey
where $w_{AB}^k(r_{p,i})$ is the value of $w_{AB}$ at the $i$-th separation bin from the $k$-th realization,
and $\overline{w}_{AB}(r_p,i) = \frac{1}{N}\sum_k w_{AB}^k(r_p,i)$.
The number of the realizations created using the Jackknife resampling is chosen to be sufficiently larger than the number of bins to obtain a non-singular matrix,
and to be small enough that the jackknifed area becomes larger than the scales of interest.
We set $N=36$ and $64$ for W2 and W3, respectively, and thus the total number of realizations is 100, which give a grid spacing of $\sim0.8$ deg ($\sim50 h^{-1}$Mpc at $z\sim1.4$). 
We have performed the analysis by adopting a different number for the Jackknife resampling, $N=64$ and $100$ for W2 and W3, respectively, and confirmed that the different choice of $N$ does not alter the covariance matrix significantly.

Equation (\ref{eq:covariance}) is known to underestimate the statistical error due to the limited number of realizations, by a factor of $(N-N_{\rm bin}-2)/(N-1)$, where $N_{\rm bin}$ is the number of the data bins used for the analysis \citep{Hartlap:2007}. Since we adopt $N_{\rm bin}=8$ as we describe in section \ref{sec:constraints} below, this factor becomes $0.909$. 
We take into account this correction factor in the following analysis. 

\subsection{Results}\label{sec:results}

Here we show the results of the GI and II correlation function measurements. 
We present the projected statistics as our main results [equation (\ref{equation:GI_proj})].
These measurements are done in the scale of $1.35<r_p<45 h^{-1}$Mpc with a binning size $\Delta \log_{10}r_p=0.2$.

\begin{figure}[bt]
 \begin{center}
 \includegraphics[width=0.45\textwidth] {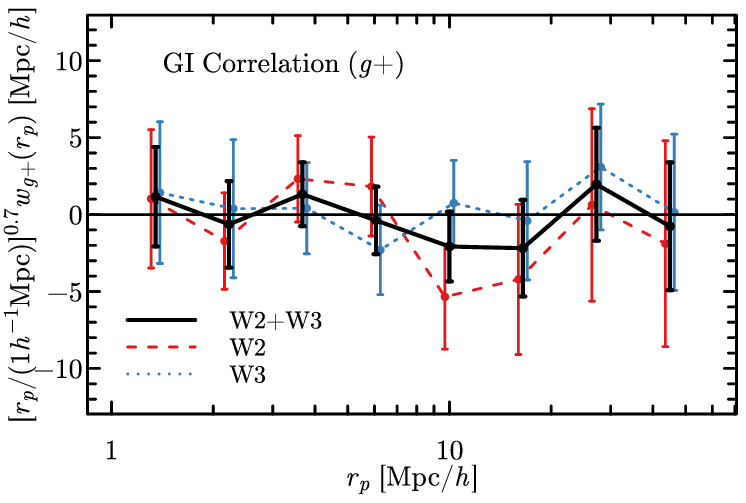}
  \includegraphics[width=0.45\textwidth]{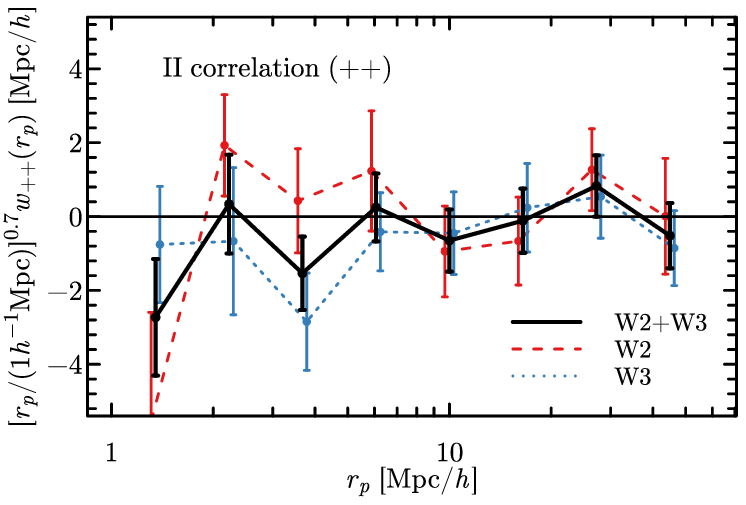}
  \includegraphics[width=0.45\textwidth]{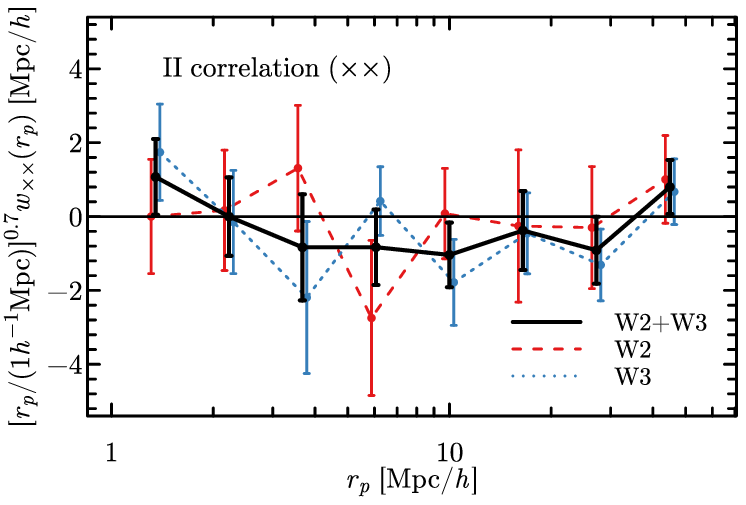}  
 \end{center}
 \caption{Projected correlation functions, $w_{g+}$ (top), $w_{++}$ (middle) and $w_{\times\times}$ (bottom) as a function of transverse separation $r_p$.
The vertical axes are multiplied by $r_p^{0.7}$ for a presentation purpose.
The red dashed and blue dotted lines are the results for the CFHTLS sub-fields, W2 and W3, respectively, while the black solid line is the combined result. The error bars are obtained from the diagonal elements of the covariance matrix, $C_{ii}^{1/2}$. 
}
\label{figure:GI+_II++_proj}
\end{figure}

In the top panel of figure \ref{figure:GI+_II++_proj}, we present the projected GI correlation $w_{g+}(r_p)$. 
The black line is our main result, the GI correlation function measured from the combined sample of W2 and W3 fields. It is fully consistent with zero within the $1-\sigma$ error, which implies that the major axes of emission line galaxies at $z\sim 1.4$ are randomly oriented. We compute the $\chi^2$ values for a fit to zero signal, for $w_{g+}(r_p)$, using the covariance matrices of these correlations, and
the reduced $\chi^2$ value is $0.29$, implying that the GI signals are indeed consistent with zero. 
We will present the more quantitative analysis by including the full covariance matrix in section \ref{sec:constraints}. 
To see the variation of the measurement in different fields, we also plot the GI functions measured from each of the W2 and W3 fields, shown as the red dashed and blue dotted points, respectively. They are all consistent with each other and with zero signal within the error bars.

Next, we consider the II correlation function. 
At the redshifts of the FastSound survey, $z\sim 1.36$, the amplitude of the II correlation is an order of magnitude smaller than that of the GI correlation in the linear alignment model (HS04). Thus, considering the null detection of the GI correlation, measuring the II correlation is useful to check  if there is any systematic effect and if the linear alignment model is really a correct model, as emphasized by \citet{Mandelbaum:2011}. 
The middle and bottom panels of Figure \ref{figure:GI+_II++_proj} show the measured II correlation functions, $w_{++}(r_p)$ and $w_{\times\times}(r_p)$, respectively.
As expected, the measurements are consistent with zero, with the reduced $\chi^2$ value of $0.86$ and $0.73$ for $w_{++}(r_p)$ and $w_{\times\times}(r_p)$ respectively.
Both the PSF distortions and intrinsic ellipticities yield positive correlations on small scales.
Therefore, our null detection of IA is in fact unlikely due to cancellation between systematic effects and the true II signal.
Further systematic tests will be presented in \S\ref{subsection:systematics}.

\begin{figure}
 \begin{center}
  \includegraphics[width=0.45\textwidth] {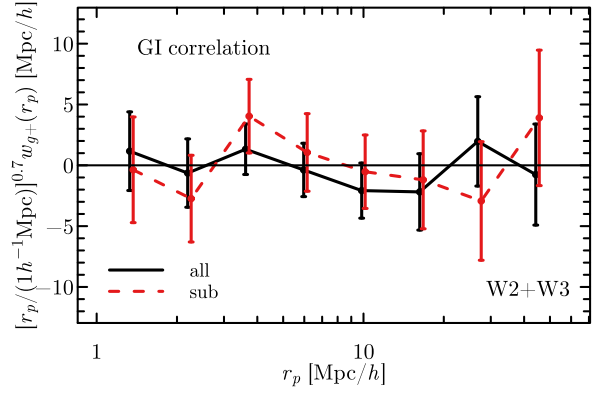}
 \end{center}
 \caption{The black points are the same as those in the top panel of figure \ref{figure:GI+_II++_proj}, the GI correlation of the shape sample with the whole FastSound galaxies. The red points are similar to the black points but we use the subsample of the FastSound galaxies which have the shape information.}
\label{figure:GIp_proj_sub}
\end{figure}

\subsection{Systematic tests}\label{subsection:systematics}
To confirm our null detection of IA, we perform various systematic tests. 
As we mentioned in section \ref{sec:data}, only the half of our FastSound galaxy data has the shape information. 
To see the possibility that IA signals are smeared out due to the stochasticity between the two populations, 
we measure the GI correlation using only the galaxies which have shapes from the CFHTLenS data. 
The result is shown as the red points in figure \ref{figure:GIp_proj_sub}. 
As expected, the size of the error bars becomes larger by $\sim 30\%$ than the case where the whole FastSound galaxy data are used, 
since the size of the shape sample is half of the whole sample. 
The correlation is still consistent with zero within the statistical scatter, indicating that 
the null detection of the GI correlation is not due to the stochasticity between the shape and whole galaxy samples.

\begin{figure}
 \begin{center}
  \includegraphics[width=0.45\textwidth] {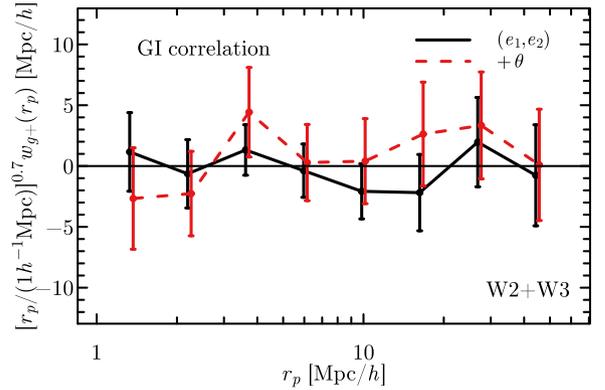}
 \end{center}
 \caption{The black points are the same as those in the top panel of figure \ref{figure:GI+_II++_proj}. 
 The red points are similar to the black points but the GI correlation with the ellipticity determined using equation (\ref{equation:eplus_theta}) instead of equation (\ref{equation:eplus}).
 }
\label{figure:GIp_proj_theta}
\end{figure}

Next, we test how our definitions of ellipticities could affect our result. 
For the main result, the ellipticity of our galaxy sample to measure the correlation function is determined 
by equation (\ref{equation:eplus}). 
The red points in figure \ref{figure:GIp_proj_theta} are the result when equation (\ref{equation:eplus_theta}) is used 
instead of (\ref{equation:eplus}), where $(e_1,e_2)$ and $\theta$ are determined by the different 
processes (see section \ref{sec:shape}). 
Although they are different ways, both the results are consistent with each other and also with zero.
Thus, we conclude our result is not sensitive to the choice of the definition for the galaxy ellipticities. 

\begin{figure}[bt]
 \begin{center}
 \includegraphics[width=0.45\textwidth] {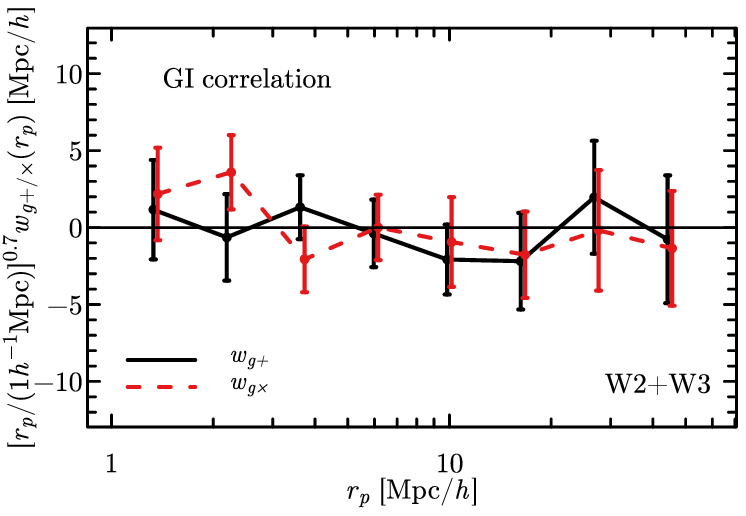}
  \includegraphics[width=0.45\textwidth] {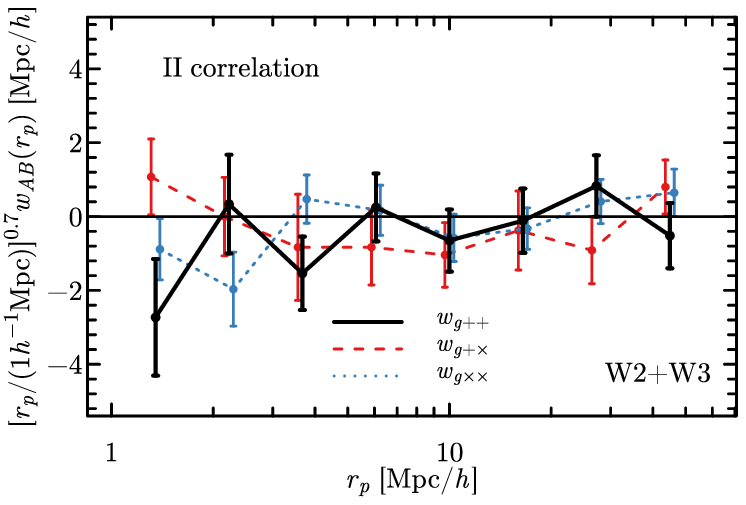}
   \end{center}
 \caption{(Top panel) The black points are $w_{g+}$, the same as those in the top panel of figure \ref{figure:GI+_II++_proj}, and the red points are $w_{g\times}$. (Bottom panel) The black points and red points are $w_{++}$ and $w_{\times\times}$, the same as the black points in the middle and bottom panels of figure \ref{figure:GI+_II++_proj}, respectively, and the blue points are the cross-correlation component $w_{+\times}$. 
 }
\label{figure:GIx_II+x_proj}
\end{figure}

Finally, we perform the so-called $45^{\circ}$ tests. 
The cross correlations, $w_{g\times}$ and $w_{+\times}$, should be zero by symmetry, and their non-zero signals can be generated only by non-physical sources.
Thus measuring these quantities provides additional checks for systematic effects such as the calibration error in the shape measurement.
The results for $w_{g\times}$ and $w_{+\times}$ are shown in the upper and lower panels of figure \ref{figure:GIx_II+x_proj}, respectively.
For comparison, the corresponding GI and II correlations are shown in each of the panels.
These statistics are found to be consistent with zero, again confirming our null detection of the IA signals. 

\subsection{Other alignment statistics} \label{subsection:alignment_statistics}
In addition to the projected statistics described above, we have considered various statistics of IA. 
We measured the monopole components of the GI and II correlation functions, respectively $\xi_{g+}(r)$ and $\xi_{++}(r)$, 
where $r=|{\bf r}|$. These quantities are not affected by the projection, while they are not significantly 
affected by RSD on large scales \citep{Okumura:2017a}. 
However, we could not see any improvement due to the large error bars and the results were consistent with zero. 

Another statistics we considered was the alignment correlation function \citep{Paz:2008,Faltenbacher:2009}. 
It is defined as the galaxy two-point correlation function as a function of not only the separation ${\bf r}$ 
but also the angle between the major axis of the galaxy and 
${\bf r}$ projected onto the sky, $\theta-\phi$, namely $\xi_{gg}({\bf r},\theta-\phi)$. 
We measured the alignment correlation for a consistency check because this statistics contains 
the information on IA almost equivalent to the GI correlation function.
We have confirmed that the trend of null-detection of IA in the alignment correlation 
is the same as that seen in the GI correlation. 

\section{Constraints on IA and cosmological implications}\label{sec:constraints}
In this section, we quantify the effect of IA of the FastSound galaxies 
by fitting the observed IA correlations with two models: a power-law model and the linear alignment model.
Results obtained using the former model are used to compare with previous studies,
while those using the latter model is used to estimate possible biases on $\sigma_8$ and $\Omega_m$ determinations from weak-lensing surveys
when the observed IA are ignored.
We assume the linear galaxy bias (equation (\ref{equation:lbias})) and use the best-fitting value of $b_g=1.9$ obtained by the RSD analysis by Paper IV.

\begin{figure}
 \begin{center}
   \includegraphics[width=0.45\textwidth] {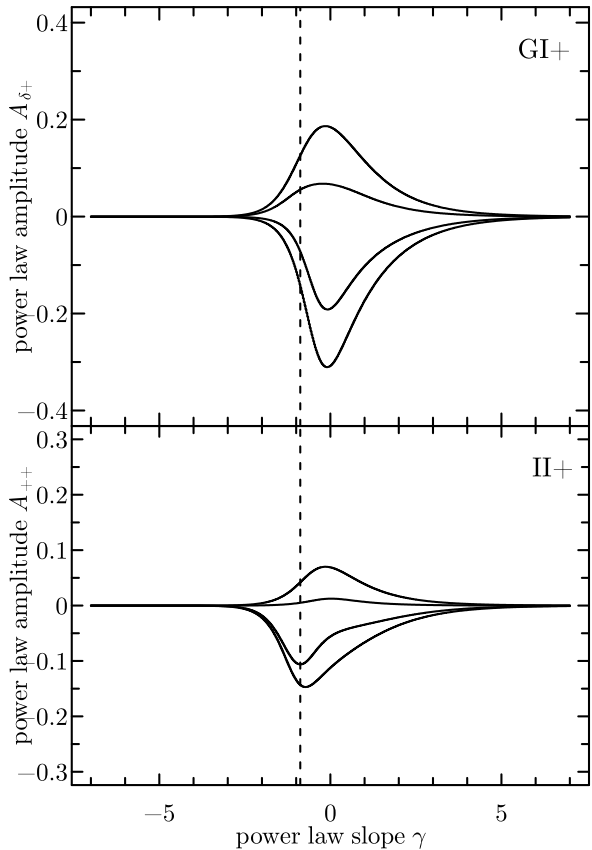}
 \end{center}
 \caption{Constraint on the fitting parameters of the power-law to the GI correlation $w_{g+}(r_p)$ (top) and the II correlation $w_{++}(r_p)$ (bottom).
 The two contours in each panel show $68\%$ and $95\%$ of the probability.
 The vertical lines indicate $\gamma_{g+}=\gamma_{++}=-0.88$. 
 }
\label{figure:constr_PL}
\end{figure}

\subsection{Power-law alignment model}
We fit the power law model denoted by tilde, 
\begin{equation}\label{equation:powerlaw}
\widetilde{w}_{\delta +}(r_p) = A_{\delta +} \left( \frac{r_p}{20h^{-1} {\rm Mpc}} \right)^{\gamma_{\delta +}}(1-f_{\rm blund})^2
\end{equation}
to the projected GI correlation function $w_{g+}(r_p)$,
where $(A_{\delta+},\gamma_{\delta +})$ are the fitting parameters and $f_{\rm blund}=0.071$ is the redshift blunder rate of our sample.
We also fit the power law to $w_{++}(r_p)$ in the same manner with $(A_{++},\gamma_{++})$.

Figure \ref{figure:constr_PL} presents the joint likelihood $L$
on the amplitude $A_{\delta +}$ and slope $\gamma_{\delta +}$ for $\wt{w}_{g+}(r_p)$,
computed by $L=\exp(-\chi^2/2)$, where $\chi^2=(w_{g+}-\wt{w}_{g+})^T C^{-1} (w_{g+}-\wt{w}_{g+})$,
and similarly for $\wt{w}_{++}(r_p)$.
As expected from the measurements of $w_{g+}$ and $w_{++}$, 
the best-fitting values of $A_{\delta+}$ and $A_{++}$ are very close to zero.
The weak constraints on $\gamma_{\delta +}$ and $\gamma_{++}$ come from the fact that the slope of the power law model cannot be determined when the amplitude is zero. 
We give constraints on $A_{\delta +}$ and $A_{++}$ by fixing $\gamma_{\delta+}=\gamma_{++}=-0.88$, the value 
adopted by the previous studies on IA \citep{Mandelbaum:2006,Mandelbaum:2011},
in order to compare the amplitudes of IA at different redshifts
in a consistent manner. We obtain the constraint on $A_{\delta +}$ as $-0.134<A_{\delta +}<0.134$ ($95\%$ confidence level)
and $-0.014<A_{++}<0.035$,
which is weaker than the that from the WiggleZ survey at $z>0.52$
due to the difference of the sample sizes.

\begin{table}
\footnotesize
\begin{center}
  \caption{The comparison of 95 per cent confidence limits of the power law amplitudes from several redshift ranges,
with the power-law slope being fixed to $\gamma=-0.88$.}
\begin{tabular}{cccc}
\hline
\hline
 Data & type & redshift & A \\
\hline
SDSS Main Blue L4 &  $w_{g+}$ & $0.09$ & $0.0160_{-0.0195}^{+0.0192}$ \\
WiggleZ, $z<0.52$ & $w_{g+}$ & $0.37$ & $0.0260_{-0.0706}^{+0.0704}$ \\
WiggleZ, $z>0.52$ & $w_{g+}$ & $0.62$ & $-0.0030_{-0.0373}^{+0.0368}$ \\
FastSound & $w_{g+}$ & $1.36$ & $-0.0071_{-0.1340}^{+0.1340}$ \\
&&&\\
SDSS Main Blue L4 &  $w_{++}$ & $0.09$ & $0.0000_{-0.0004}^{+0.0008}$ \\
WiggleZ, $z<0.52$ & $w_{++}$ & $0.37$ & $-0.0130_{-0.0254}^{+0.0250}$ \\
WiggleZ, $z>0.52$ & $w_{++}$ & $0.62$ & $0.0125_{-0.0209}^{+0.0210}$ \\
FastSound & $w_{++}$ & $1.36$ & $-0.0505_{-0.0858}^{+0.0858}$ \\
\hline
\end{tabular}
\label{table:PL}
\end{center}
\end{table}

\begin{figure}
 \begin{center}
 \includegraphics[width=0.45\textwidth] {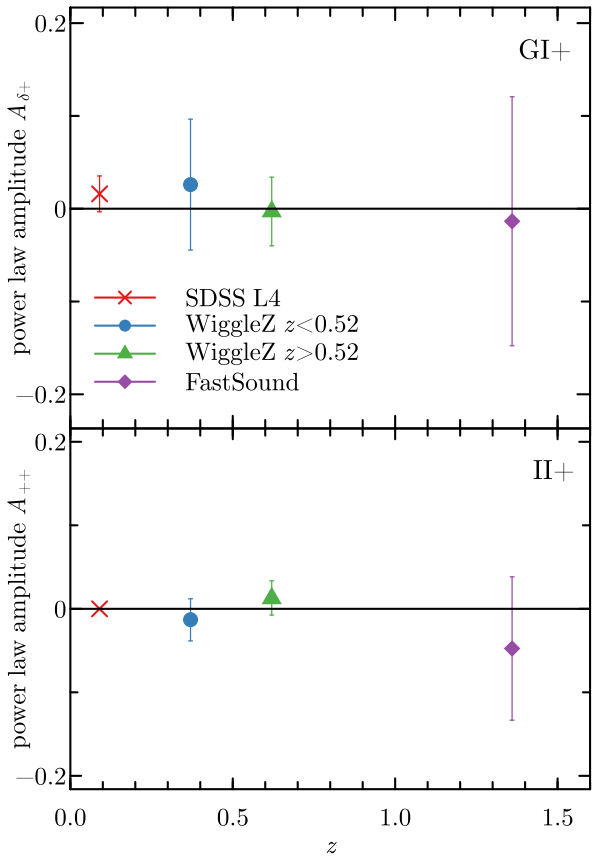}
 \end{center}
 \caption{Constraints on the amplitude of the power-law model for IA of star-forming galaxies, $A_{g+}$ (top) and $A_{++}$ (bottom) as a function of redshift. 
 The constraint obtained from our FastSound sample at $1.19<z<1.55$ is plotted as the purple point. The previous results at lower redshifts
 from the SDSS and WiggleZ surveys are also plotted. 
 }
\label{figure:z_PLamp}
\end{figure}

Table \ref{table:PL} and figure \ref{figure:z_PLamp} summarize the constraints on the IA amplitudes $A$ from our data,
together with the previous studies on blue galaxies \citep{Mandelbaum:2006,Mandelbaum:2011}.
Let us address here the similarity and difference of the FastSound sample compared to the samples used in low-$z$ studies.
Both the WiggleZ and FastSound surveys targeted starburst galaxies with strong emission lines, although they were selected in different ways.
The samples in these surveys have the stellar masses similar to each other with a median value of $\sim10^{10} \rm{M_\odot}$ (\cite{Banerji:2013}; Paper III).
According to a halo occupation distribution modeling
applied to the measured galaxy clustering,
the constrained halo masses are also in good agreement (\cite{Koda:2016}; Paper IV) and most of the galaxies in these samples are central galaxies.
Therefore, these two samples comprise similar types of galaxies residing in similar environments.
On the other hand, as pointed out by \citet{Mandelbaum:2011}, the SDSS L4 blue and WiggleZ samples may have different formation histories because of their different color distributions.
With these points in mind, the null detection of IA at redshift of $1.4$ 
together with that at lower redshifts may imply that IA do not exist for blue galaxies up to $z > 1$.
This means that physical processes such as galaxy mergers and interactions, which tend to erase the alignment,
might be effective for the late-type galaxies.

Recently, the first year results from the DES survey have been released (\cite{Troxel:2017}).
They obtained constraints on cosmological parameters and IA simultaneously
by fitting the observed shear correlation function using a model which includes IA contributions 
as well as the pure lensing contribution (\cite{Blazek:2017}).
They reported the detection of the GI signal with a positive amplitude of the tidal alignment ($A_1$) component
and a negative amplitude of the tidal torquing alignment ($A_2$) component (\cite{Troxel:2017}), which is in contrast to our null detection.
It is, however, difficult to make a detailed comparison because the methods used in the DES analysis and ours are different in the following three aspects:
(1) they used the photometric redshifts rather than spectroscopic ones,
(2) they did not directly measure the correlation between galaxy shapes and the 3 dimensional matter distribution,
and (3) they did not differentiate blue/spiral and red/elliptical galaxies in the analysis and
it is not clear which galaxy type contributes to each of the $A_1$ and $A_2$ signals.

The amplitude of IA strongly depends on the host halo mass \citep{Jing:2002,van-Uitert:2017,Xia:2017,Okumura:2017a}.
The analysis of Paper IV found that the galaxies in our sample are so sparse that the majority of them are central galaxies residing in low-mass halos
(see \cite{Kashino:2017a} for the result for the denser sample).
This could be the reason for the null detection of IA in our analysis, and emission line galaxies in different environments
such as satellite galaxies in clusters and galaxies along filaments may show significant IA signals.
However, our null detection of IA could be just due to the large error bars coming from
both the limited size and number density of the FastSound survey.
This kind of possibilities can be tested by larger, ongoing and future surveys, such as the PFS, 
which span wider redshift ranges (up to $z\sim2.4$).
These data will allow for precise measurements
and more detailed studies of IA, including 
the dependences on galaxy classes, environments, and redshifts.

\subsection{Linear alignment model}
Another model that we attempt to use to analyze the measurement is the linear alignment (LA) model
(\cite{Catelan:2001}; HS04).
In this model, the cross-power spectrum between the matter density fluctuation and the galaxy-density weighted intrinsic shear,
  $P_{\delta, \tilde \gamma^{\rm I}}(k)$ (equation (\ref{eq:c_gi})), is related to the linear matter power spectrum $P^{\rm lin}_\delta(k)$ through:
\begin{equation}
P_{\delta, \tilde \gamma^{\rm I}}(k) = {C_1\bar\rho\over\bar D}a^2 P^{\rm lin}_\delta(k). \label{eq:p_la}
\end{equation}
where $C_1$ is a normalization factor, $\bar\rho(z)$ is the mean matter density,
and $\bar D\equiv (1+z)D(z)$ is the growth factor normalized to unity at the matter dominant epoch.
The normalization factor $C_1$ is an order of $5\times10^{-14}{(h^2 {\rm M_\odot Mpc^{-3}})}^{-1}$ 
at low redshift (\cite{Bridle:2007}).
Note that this equation is taken from \citet{Hirata:2010} (hereafter HS10) and is different from the original expression of HS04 by the factor of $a^2$.
We will use the corrected one of HS10 for deriving main results, while 
we also show the result based on the previous version of HS04 to compare with previous studies, 
most of which had used the HS04 model.
Also, note that
the tidal torquing model (\cite{Catelan:2001}) may give a better description for disc-like galaxies than the LA model.
However, as argued by \citet{Krause:2016}, both the linear alignment and the tidal torquing can contribute to IA of blue galaxies.
While our null detection of the II signal could be useful to rule out the quadratic model which predicts non-zero II,
it is difficult at this stage to determine the best IA model for blue galaxies given the relatively large error bars.
Therefore we use the LA model in this work considering that this model has been widely used in literature and is simple to treat.

In practice, in our analysis we use a modified version of the LA model, which replaces $P_\delta^{\rm lin}(k)$ in equation (\ref{eq:p_la}) by the non-linear matter power spectrum $P^{\rm nl}_\delta(k)$ \citep{Bridle:2007,Hirata:2007},
to incorporate the non-linear effects on the large-scale structure power spectrum:
\begin{equation}
P^{\rm nl}_{\delta, \tilde \gamma^{\rm I}}(k) = {C_1\bar\rho\over\bar D} a^2
P^{\rm nl}_\delta(k). \label{eq:p_la_nl}
\end{equation}
We calculate $P^{\rm nl}_\delta(k; z=1.4)$ using the halofit model \citep{Smith:2003} whose fitting parameters are improved by 
high-resolution $N$-body simulations by \citet{Takahashi:2012}.
The power spectrum $P^{\rm nl}_{\delta {\rm I}}(k)$ is then converted into the projected correlation function
using equation \eqref{equation:wd+}, and it can be compared to our measurements by assuming the linear bias relation (equation (\ref{equation:lbias})).
Unlike the power-law model, there is only one parameter, the amplitude $C_1$. 
We vary it and calculate the $\chi^2$ statistics, and fit the observed projected GI correlation (figure \ref{figure:GI+_II++_proj}).

\begin{figure}
 \begin{center}
  \includegraphics[width=0.45\textwidth] {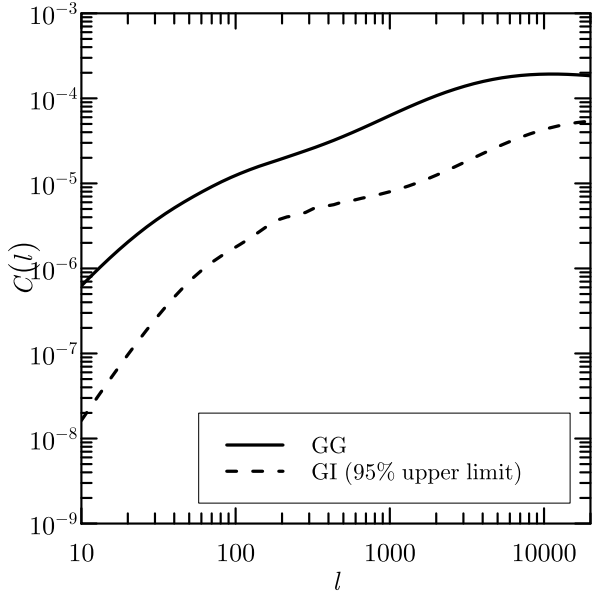}
 \end{center}
 \caption{Forecast of the projected angular power spectrum for GG and GI terms.
 The GG power spectrum is calculated with source redshift distribution taken from the FastSound spectroscopic sample, assuming $\sigma_8=0.8$.
 The GI power spectrum is based on the LA model with the amplitude $C_1/(5\times10^{-14}{(h^2 {\rm M_\odot Mpc^{-3})^{-1}})}=-17.4$.}
\label{figure:GG_GI}
\end{figure}

The best-fitting modified LA model of HS10 (equation (\ref{eq:p_la_nl})) provides
$C_1/(5\times10^{-14}{(h^2 {\rm M_\odot Mpc^{-3}})^{-1})} = 1.58^{+20.21}_{-20.21}$ ($95\%$ confidence level) for the FastSound IA,
and $0.49^{+3.56}_{-3.56}$ if we use the HS04 model (equation (\ref{eq:p_la})).
The maximum contribution of the GI angular power spectrum $C_{\rm GI}(l)$ within the $95\%$ confidence level is presented in Figure \ref{figure:GG_GI},
together with the shear power spectrum $C_{\rm GG}(l)$ for comparison,
which is obtained from equation \eqref{eq:c_gg} with a single tomographic bin
assuming the redshift distribution of the FastSound sample (Paper IV).
The contamination of the GI correlation to the total power, $C_{\rm GI}/C_{\rm GG}$, is about $\sim 15\%$ at $l\sim500$,
suggesting that the ongoing larger lensing surveys such as DES \citep{The-Dark-Energy-Survey-Collaboration:2005} and HSC \citep{Miyazaki:2012},
aiming to measure the shear power spectrum with a few percent precision,
may need to take into account the systematic error due to the the IA contaminations properly. 

Finally, we estimate how the measured GI correlation could affect the cosmological parameter estimations 
in weak lensing surveys if it was ignored. 
Although the GI signal for our sample is consistent with zero,
it is still meaningful to place the upper and lower bounds on the bias on cosmological parameters. 
We use a Fisher matrix approach (\cite{Huterer:2006}; \cite{Kirk:2015}), and 
the bias on $\alpha$-th parameter $\Delta p_{\alpha}$ is evaluated as:
\begin{eqnarray}
\begin{split}
\Delta p_{\alpha} = \sum_{\beta} F_{\alpha\beta}^{-1} \sum_{\ell} &\sum_{i\leq j; m \leq n} & \\
 \Delta C_{\rm GG}^{(ij)} (\ell) 
 \bigl( {\rm Cov } & \left[ C_{\rm GG}^{(ij)}(\ell),C_{\rm GG}^{(mn)}(\ell) \right] \bigr)^{-1} \frac{\partial C_{\rm GG}^{(mn)}(\ell)}{\partial p_{\beta}},
\label{eqn:biascl}
\end{split}
\end{eqnarray}\\
where $C_{\rm GG}^{(ij)}(\ell)$ is the cross-power spectrum between $i$-th and $j$-th
tomographic redshift bins, and $\Delta C_{\rm GG}^{(ij)}(\ell)$ is the bias on $C_{\rm GG}^{(ij)}(\ell)$.
In this analysis, we only consider the GI term as a source of bias.
The Fisher matrix $F_{\alpha\beta}$ is calculated as: 
\begin{eqnarray}
\begin{split}
F_{\alpha\beta} = \sum_{\ell} \sum_{i\leq j; m \leq n} & \\
\frac{\partial C_{\rm GG}^{(ij)}(\ell)}{\partial p_{\alpha}}   \bigl({\rm Cov } & \left[ 
C_{\rm GG}^{(ij)}(\ell),C_{\rm GG}^{(mn)}(\ell) \right] \bigr) ^{-1} \frac{\partial C_{\rm GG}^{(mn)}(\ell)}{\partial p_{\beta}}.
\label{eqn:deltacl}
\end{split}
\end{eqnarray}
Here we limit our calculation to a single redshift bin without tomography for simplicity.
Also, we only calculate the bias on each of the parameters $\sigma_8$ and $\Omega_m$, with
another being fixed to the fiducial value.
We use the observed redshift distribution of our FastSound sample to calculate $C_{\rm GG}^{(mn)}$.
The range of $\ell$ used for fitting is $100 \leq \ell \leq 5000$.

The range of bias on $\sigma_8$, assuming that only GI term contaminates the shear power spectrum,
is $-0.052<\Delta \sigma_8<0.039$ ($95\%$ limits).
We perform the same analysis for $\Omega_m$ and derive the bias as $-0.039<\Delta \Omega_m<0.030$.
These constraints do not depend on whether we use the model of HS04 or HS10, because the factor of $a^2$ is cancelled out when we derive the best-fitting value of $C_1$.
While the error bars are large, the contamination of IA into the GG correlation may reach up to $\sim10\%$,
hence the careful modeling of the GI contamination will be important for
the shear power spectrum analysis.
These values are similar to those adopted by \citet{Krause:2016}.
In figure 3 of \citet{Krause:2016}, they forecast the impacts of IA of blue galaxies in the LSST,
based on the LA model with the upper limit from the observation of \citet{Mandelbaum:2011}.
One can find $|\Delta \sigma_8| \sim0.03$ and $|\Delta \Omega_m| \sim 0.02$
by comparing the blue and red lines in the figure.
\citet{Blazek:2017} also give similar forecasts
based on their perturbative IA model, though the amplitude of the tidal torquing component is not determined from observations.
More observations will be needed to evaluate the effect of IA more precisely,
by determining the best model and amplitudes.
Also, we have limited ourselves to the case of a single redshift bin without tomography.
If a larger spectroscopic data set becomes available,
it will be possible to investigate the effect of IA even for the tomography survey case.

\section{Conclusions}\label{sec:conclusion}
In this paper, we have studied intrinsic alignments (IA) of star-forming galaxies 
by measuring the correlation between the overdensity and ellipticity of galaxies
using the spectroscopic galaxy sample of the FastSound survey and
the shape sample measured by the CFHTLenS survey.
By performing the analysis at $z\sim 1.4$,
we examined whether the non-detection of IA in the earlier studies at the intermediate redshift,  $z\sim 0.7$, 
is expanded to such a high redshift, 
which can give an insight on how physical processes
such as galaxy mergers affect the evolution of galaxy shapes.
We also studied the extent to which IA contaminate the signal
of the weak lensing power spectrum and change the inferred cosmological parameters.

We measured the II and GI correlation functions and fitted them by a power-law model of IA following the previous studies. 
We then found that the correlations are entirely consistent with zero within the error bars. 
Combined with the previous results up to the intermediate redshift, 
the IA signal for blue galaxies does not exist over a wide range of redshifts.

We also used the linear alignment (LA) model with the non-linear matter power spectrum
for fitting the GI correlation to determine the upper limit of IA and possible contamination to cosmological parameter estimations in the weak lensing analysis.
Using the $95\%$ confidence interval of this amplitude,
we showed that the maximum contamination from the GI correlation to the weak lensing signal is up to a few percent,
if we assume the same redshift distribution of source galaxies as our sample.
From a Fisher analysis, this contamination is converted into the bias on cosmological parameter estimates,
$-0.052<\Delta \sigma_8<0.039$ and $-0.039<\Delta \Omega_m<0.030$, if we choose the fiducial values of $\Omega_m=0.3$ and $\sigma_8=0.8$, respectively.

Since our galaxy sample is not large, the measurement of IA is noisy. 
Ongoing and future galaxy surveys such as PFS will enable us to improve the accuracies. 
It will allow for more detailed studies of IA as a function of galaxy type, environment, and redshift,
which will be useful both for the study of cosmology and galaxy formation/evolution.

\bigskip

We thank Tomotsugu Goto for helpful comments on the manuscript.
We are also grateful to the anonymous referee for useful comments.
The FastSound project was supported in part by MEXT/JSPS KAKENHI Grant Numbers 19740099, 19035005, 20040005, 22012005, and 23684007. 
This work is in part based on data collected at Subaru Telescope, which is operated by the National Astronomical Observatory of Japan.
T. O. acknowledges support from the Ministry of Science and Technology of Taiwan under the grant MOST 106-2119-M-001-031-MY3.

\bibliography{ms_ver7.0.bbl}

\begin{thebibliography}{}
\expandafter\ifx\csname natexlab\endcsname\relax\def\natexlab#1{#1}\fi

\bibitem[{{Banerji} {et~al.}(2013){Banerji}, {Glazebrook}, {Blake}, {Brough},
  {Colless}, {Contreras}, {Couch}, {Croton}, {Croom}, {Davis}, {Drinkwater},
  {Forster}, {Gilbank}, {Gladders}, {Jelliffe}, {Jurek}, {Li}, {Madore},
  {Martin}, {Pimbblet}, {Poole}, {Pracy}, {Sharp}, {Wisnioski}, {Woods},
  {Wyder}, \& {Yee}}]{Banerji:2013}
{Banerji}, M., {Glazebrook}, K., {Blake}, C., {et~al.} 2013, \mnras, 431, 2209

\bibitem[{{Bartelmann} \& {Schneider}(2001)}]{Bartelmann:2001}
{Bartelmann}, M., \& {Schneider}, P. 2001, \physrep, 340, 291

\bibitem[{{Benson}(2010)}]{Benson:2010}
{Benson}, A.~J. 2010, \physrep, 495, 33

\bibitem[Bertin \& Arnouts(1996)]{Bertin:1996} Bertin, E., \& Arnouts, S.\ 1996, \aaps, 117, 393 

\bibitem[{{Binney}(1977)}]{Binney:1977}
{Binney}, J. 1977, \apj, 215, 483

\bibitem[{{Binney} \& {de Vaucouleurs}(1981)}]{Binney:1981}
{Binney}, J., \& {de Vaucouleurs}, G. 1981, \mnras, 194, 679

\bibitem[{{Blazek} {et~al.}(2017){Blazek}, {MacCrann}, {Troxel}, \&
  {Fang}}]{Blazek:2017}
{Blazek}, J., {MacCrann}, N., {Troxel}, M.~A., \& {Fang}, X. 2017, ArXiv
  e-prints, arXiv:1708.09247

\bibitem[{{Blazek} {et~al.}(2011){Blazek}, {McQuinn}, \&
  {Seljak}}]{Blazek:2011}
{Blazek}, J., {McQuinn}, M., \& {Seljak}, U. 2011, \jcap, 5, 10

\bibitem[{{Bridle} \& {King}(2007)}]{Bridle:2007}
{Bridle}, S., \& {King}, L. 2007, New Journal of Physics, 9, 444

\bibitem[{{Catelan} {et~al.}(2001){Catelan}, {Kamionkowski}, \&
  {Blandford}}]{Catelan:2001}
{Catelan}, P., {Kamionkowski}, M., \& {Blandford}, R.~D. 2001, \mnras, 320, L7

\bibitem[{{Chisari} {et~al.}(2015){Chisari}, {Codis}, {Laigle}, {Dubois},
  {Pichon}, {Devriendt}, {Slyz}, {Miller}, {Gavazzi}, \&
  {Benabed}}]{Chisari:2015}
{Chisari}, N., {Codis}, S., {Laigle}, C., {et~al.} 2015, \mnras, 454, 2736

\bibitem[{{Ciotti} \& {Dutta}(1994)}]{Ciotti:1994}
{Ciotti}, L., \& {Dutta}, S.~N. 1994, \mnras, 270, 390

\bibitem[{{Croft} \& {Metzler}(2000)}]{Croft:2000}
{Croft}, R.~A.~C., \& {Metzler}, C.~A. 2000, \apj, 545, 561

\bibitem[{{Dubois} {et~al.}(2016){Dubois}, {Peirani}, {Pichon}, {Devriendt},
  {Gavazzi}, {Welker}, \& {Volonteri}}]{Dubois:2016}
{Dubois}, Y., {Peirani}, S., {Pichon}, C., {et~al.} 2016, \mnras, 463, 3948

\bibitem[{{Erben} {et~al.}(2013){Erben}, {Hildebrandt}, {Miller}, {van
  Waerbeke}, {Heymans}, {Hoekstra}, {Kitching}, {Mellier}, {Benjamin}, {Blake},
  {Bonnett}, {Cordes}, {Coupon}, {Fu}, {Gavazzi}, {Gillis}, {Grocutt}, {Gwyn},
  {Holhjem}, {Hudson}, {Kilbinger}, {Kuijken}, {Milkeraitis}, {Rowe},
  {Schrabback}, {Semboloni}, {Simon}, {Smit}, {Toader}, {Vafaei}, {van Uitert},
  \& {Velander}}]{Erben:2013}
{Erben}, T., {Hildebrandt}, H., {Miller}, L., {et~al.} 2013, \mnras, 433, 2545

\bibitem[{{Faltenbacher} {et~al.}(2009){Faltenbacher}, {Li}, {White}, {Jing},
  {Mao}, \& {Wang}}]{Faltenbacher:2009}
{Faltenbacher}, A., {Li}, C., {White}, S.~D.~M., {et~al.} 2009, Research in
  Astronomy and Astrophysics, 9, 41

\bibitem[{{Gwyn}(2012)}]{Gwyn:2012}
{Gwyn}, S.~D.~J. 2012, \aj, 143, 38

\bibitem[{{Hartlap} {et~al.}(2007){Hartlap}, {Simon}, \&
  {Schneider}}]{Hartlap:2007}
{Hartlap}, J., {Simon}, P., \& {Schneider}, P. 2007, \aap, 464, 399

\bibitem[{{Heavens} {et~al.}(2000){Heavens}, {Refregier}, \&
  {Heymans}}]{Heavens:2000}
{Heavens}, A., {Refregier}, A., \& {Heymans}, C. 2000, \mnras, 319, 649

\bibitem[{{Heymans} {et~al.}(2004){Heymans}, {Brown}, {Heavens},
  {Meisenheimer}, {Taylor}, \& {Wolf}}]{Heymans:2004}
{Heymans}, C., {Brown}, M., {Heavens}, A., {et~al.} 2004, \mnras, 347, 895

\bibitem[{{Heymans} {et~al.}(2012){Heymans}, {Van Waerbeke}, {Miller}, {Erben},
  {Hildebrandt}, {Hoekstra}, {Kitching}, {Mellier}, {Simon}, {Bonnett},
  {Coupon}, {Fu}, {Harnois D{\'e}raps}, {Hudson}, {Kilbinger}, {Kuijken},
  {Rowe}, {Schrabback}, {Semboloni}, {van Uitert}, {Vafaei}, \&
  {Velander}}]{Heymans:2012}
{Heymans}, C., {Van Waerbeke}, L., {Miller}, L., {et~al.} 2012, \mnras, 427,
  146

\bibitem[{{Hirata} {et~al.}(2007){Hirata}, {Mandelbaum}, {Ishak}, {Seljak},
  {Nichol}, {Pimbblet}, {Ross}, \& {Wake}}]{Hirata:2007}
{Hirata}, C.~M., {Mandelbaum}, R., {Ishak}, M., {et~al.} 2007, \mnras, 381,
  1197

\bibitem[{{Hirata} \& {Seljak}(2004)}]{Hirata:2004}
{Hirata}, C.~M., \& {Seljak}, U. 2004, \prd, 70, 063526

\bibitem[{{Hirata} \& {Seljak}(2010)}]{Hirata:2010}
---. 2010, \prd, 82, 049901

\bibitem[{{Huterer} {et~al.}(2006){Huterer}, {Takada}, {Bernstein}, \&
  {Jain}}]{Huterer:2006}
{Huterer}, D., {Takada}, M., {Bernstein}, G., \& {Jain}, B. 2006, \mnras, 366,
  101

\bibitem[{{Iwamuro} {et~al.}(2012){Iwamuro}, {Moritani}, {Yabe}, {Sumiyoshi},
  {Kawate}, {Tamura}, {Akiyama}, {Kimura}, {Takato}, {Tait}, {Ohta}, {Totani},
  {Suzuki}, \& {Tonegawa}}]{Iwamuro:2012}
{Iwamuro}, F., {Moritani}, Y., {Yabe}, K., {et~al.} 2012, \pasj, 64, 59

\bibitem[{{Jing}(2002)}]{Jing:2002}
{Jing}, Y.~P. 2002, \mnras, 335, L89

\bibitem[{{Joachimi} {et~al.}(2011){Joachimi}, {Mandelbaum}, {Abdalla}, \&
  {Bridle}}]{Joachimi:2011}
{Joachimi}, B., {Mandelbaum}, R., {Abdalla}, F.~B., \& {Bridle}, S.~L. 2011,
  \aap, 527, A26

\bibitem[{{Joachimi} {et~al.}(2015){Joachimi}, {Cacciato}, {Kitching},
  {Leonard}, {Mandelbaum}, {Sch{\"a}fer}, {Sif{\'o}n}, {Hoekstra}, {Kiessling},
  {Kirk}, \& {Rassat}}]{Joachimi:2015}
{Joachimi}, B., {Cacciato}, M., {Kitching}, T.~D., {et~al.} 2015, \ssr, 193, 1

\bibitem[{{Kaiser}(1987)}]{Kaiser:1987}
{Kaiser}, N. 1987, \mnras, 227, 1

\bibitem[{{Kashino} {et~al.}(2017){Kashino}, {More}, {Silverman}, {Daddi},
  {Renzini}, {Sanders}, {Rodighiero}, {Puglisi}, {Kajisawa}, {Valentino},
  {Kartaltepe}, {Le F{\`e}vre}, {Nagao}, {Arimoto}, \&
  {Sugiyama}}]{Kashino:2017a}
{Kashino}, D., {More}, S., {Silverman}, J.~D., {et~al.} 2017, \apj, 843, 138

\bibitem[{{Kilbinger} {et~al.}(2013){Kilbinger}, {Fu}, {Heymans}, {Simpson},
  {Benjamin}, {Erben}, {Harnois-D{\'e}raps}, {Hoekstra}, {Hildebrandt},
  {Kitching}, {Mellier}, {Miller}, {Van Waerbeke}, {Benabed}, {Bonnett},
  {Coupon}, {Hudson}, {Kuijken}, {Rowe}, {Schrabback}, {Semboloni}, {Vafaei},
  \& {Velander}}]{Kilbinger:2013}
{Kilbinger}, M., {Fu}, L., {Heymans}, C., {et~al.} 2013, \mnras, 430, 2200

\bibitem[{{Kimura} {et~al.}(2010){Kimura}, {Maihara}, {Iwamuro}, {Akiyama},
  {Tamura}, {Dalton}, {Takato}, {Tait}, {Ohta}, {Eto}, {Mochida}, {Elms},
  {Kawate}, {Kurakami}, {Moritani}, {Noumaru}, {Ohshima}, {Sumiyoshi}, {Yabe},
  {Brzeski}, {Farrell}, {Frost}, {Gillingham}, {Haynes}, {Moore}, {Muller},
  {Smedley}, {Smith}, {Bonfield}, {Brooks}, {Holmes}, {Curtis Lake}, {Lee},
  {Lewis}, {Froud}, {Tosh}, {Woodhouse}, {Blackburn}, {Content}, {Dipper},
  {Murray}, {Sharples}, \& {Robertson}}]{Kimura:2010}
{Kimura}, M., {Maihara}, T., {Iwamuro}, F., {et~al.} 2010, \pasj, 62, 1135

\bibitem[{{King} \& {Schneider}(2002)}]{King:2002}
{King}, L., \& {Schneider}, P. 2002, \aap, 396, 411

\bibitem[{{Kirk} {et~al.}(2015){Kirk}, {Brown}, {Hoekstra}, {Joachimi},
  {Kitching}, {Mandelbaum}, {Sif{\'o}n}, {Cacciato}, {Choi}, {Kiessling},
  {Leonard}, {Rassat}, \& {Sch{\"a}fer}}]{Kirk:2015}
{Kirk}, D., {Brown}, M.~L., {Hoekstra}, H., {et~al.} 2015, \ssr, 193, 139

\bibitem[{{Koda} {et~al.}(2016){Koda}, {Blake}, {Beutler}, {Kazin}, \&
  {Marin}}]{Koda:2016}
{Koda}, J., {Blake}, C., {Beutler}, F., {Kazin}, E., \& {Marin}, F. 2016,
  \mnras, 459, 2118

\bibitem[{{Krause} {et~al.}(2016){Krause}, {Eifler}, \& {Blazek}}]{Krause:2016}
{Krause}, E., {Eifler}, T., \& {Blazek}, J. 2016, \mnras, 456, 207

\bibitem[{{Kuhlen} {et~al.}(2007){Kuhlen}, {Diemand}, \& {Madau}}]{Kuhlen:2007}
{Kuhlen}, M., {Diemand}, J., \& {Madau}, P. 2007, \apj, 671, 1135

\bibitem[{{Lambas} {et~al.}(1992){Lambas}, {Maddox}, \&
  {Loveday}}]{Lambas:1992}
{Lambas}, D.~G., {Maddox}, S.~J., \& {Loveday}, J. 1992, \mnras, 258, 404

\bibitem[{{Landy} \& {Szalay}(1993)}]{Landy:1993}
{Landy}, S.~D., \& {Szalay}, A.~S. 1993, \apj, 412, 64

\bibitem[{{Lawrence} {et~al.}(2007){Lawrence}, {Warren}, {Almaini}, {Edge},
  {Hambly}, {Jameson}, {Lucas}, {Casali}, {Adamson}, {Dye}, {Emerson},
  {Foucaud}, {Hewett}, {Hirst}, {Hodgkin}, {Irwin}, {Lodieu}, {McMahon},
  {Simpson}, {Smail}, {Mortlock}, \& {Folger}}]{Lawrence:2007}
{Lawrence}, A., {Warren}, S.~J., {Almaini}, O., {et~al.} 2007, \mnras, 379,
  1599

\bibitem[{{LSST Dark Energy Science
  Collaboration}(2012)}]{LSST-Dark-Energy-Science-Collaboration:2012}
{LSST Dark Energy Science Collaboration}. 2012, ArXiv e-prints, arXiv:1211.0310

\bibitem[{{Mandelbaum} {et~al.}(2006){Mandelbaum}, {Hirata}, {Ishak}, {Seljak},
  \& {Brinkmann}}]{Mandelbaum:2006}
{Mandelbaum}, R., {Hirata}, C.~M., {Ishak}, M., {Seljak}, U., \& {Brinkmann},
  J. 2006, \mnras, 367, 611

\bibitem[{{Mandelbaum} {et~al.}(2011){Mandelbaum}, {Blake}, {Bridle},
  {Abdalla}, {Brough}, {Colless}, {Couch}, {Croom}, {Davis}, {Drinkwater},
  {Forster}, {Glazebrook}, {Jelliffe}, {Jurek}, {Li}, {Madore}, {Martin},
  {Pimbblet}, {Poole}, {Pracy}, {Sharp}, {Wisnioski}, {Woods}, \&
  {Wyder}}]{Mandelbaum:2011}
{Mandelbaum}, R., {Blake}, C., {Bridle}, S., {et~al.} 2011, \mnras, 410, 844

\bibitem[Miller et al.(2007)]{Miller:2007} Miller, L., Kitching, T.~D.,
Heymans, C., Heavens, A.~F., \& van Waerbeke, L.\ 2007, \mnras, 382, 315 

\bibitem[{{Miller} {et~al.}(2013){Miller}, {Heymans}, {Kitching}, {van
  Waerbeke}, {Erben}, {Hildebrandt}, {Hoekstra}, {Mellier}, {Rowe}, {Coupon},
  {Dietrich}, {Fu}, {Harnois-D{\'e}raps}, {Hudson}, {Kilbinger}, {Kuijken},
  {Schrabback}, {Semboloni}, {Vafaei}, \& {Velander}}]{Miller:2013}
{Miller}, L., {Heymans}, C., {Kitching}, T.~D., {et~al.} 2013, \mnras, 429,
  2858

\bibitem[{{Miyazaki} {et~al.}(2012){Miyazaki}, {Komiyama}, {Nakaya}, {Kamata},
  {Doi}, {Hamana}, {Karoji}, {Furusawa}, {Kawanomoto}, {Morokuma}, {Ishizuka},
  {Nariai}, {Tanaka}, {Uraguchi}, {Utsumi}, {Obuchi}, {Okura}, {Oguri},
  {Takata}, {Tomono}, {Kurakami}, {Namikawa}, {Usuda}, {Yamanoi}, {Terai},
  {Uekiyo}, {Yamada}, {Koike}, {Aihara}, {Fujimori}, {Mineo}, {Miyatake},
  {Yasuda}, {Nishizawa}, {Saito}, {Tanaka}, {Uchida}, {Katayama}, {Wang},
  {Chen}, {Lupton}, {Loomis}, {Bickerton}, {Price}, {Gunn}, {Suzuki},
  {Miyazaki}, {Muramatsu}, {Yamamoto}, {Endo}, {Ezaki}, {Itoh}, {Miwa},
  {Yokota}, {Matsuda}, {Ebinuma}, \& {Takeshi}}]{Miyazaki:2012}
{Miyazaki}, S., {Komiyama}, Y., {Nakaya}, H., {et~al.} 2012, in \procspie, Vol.
  8446, Ground-based and Airborne Instrumentation for Astronomy IV, 84460Z

\bibitem[{{Okada} {et~al.}(2016){Okada}, {Totani}, {Tonegawa}, {Akiyama},
  {Dalton}, {Glazebrook}, {Iwamuro}, {Ohta}, {Takato}, {Tamura}, {Yabe},
  {Bunker}, {Goto}, {Hikage}, {Ishikawa}, {Okumura}, \& {Shimizu}}]{Okada:2016}
{Okada}, H., {Totani}, T., {Tonegawa}, M., {et~al.} 2016, \pasj, 68, 47

\bibitem[{{Okamoto} {et~al.}(2005){Okamoto}, {Eke}, {Frenk}, \&
  {Jenkins}}]{Okamoto:2005}
{Okamoto}, T., {Eke}, V.~R., {Frenk}, C.~S., \& {Jenkins}, A. 2005, \mnras,
  363, 1299

\bibitem[{{Okumura} \& {Jing}(2009)}]{Okumura:2009a}
{Okumura}, T., \& {Jing}, Y.~P. 2009, \apjl, 694, L83

\bibitem[{{Okumura} {et~al.}(2009){Okumura}, {Jing}, \& {Li}}]{Okumura:2009}
{Okumura}, T., {Jing}, Y.~P., \& {Li}, C. 2009, \apj, 694, 214

\bibitem[{{Okumura} {et~al.}(2017){Okumura}, {Nishimichi}, {Umetsu}, \&
  {Osato}}]{Okumura:2017a}
{Okumura}, T., {Nishimichi}, T., {Umetsu}, K., \& {Osato}, K. 2017, ArXiv
  e-prints, arXiv:1706.08860

\bibitem[{{Okumura} {et~al.}(2016){Okumura}, {Hikage}, {Totani}, {Tonegawa},
  {Okada}, {Glazebrook}, {Blake}, {Ferreira}, {More}, {Taruya}, {Tsujikawa},
  {Akiyama}, {Dalton}, {Goto}, {Ishikawa}, {Iwamuro}, {Matsubara},
  {Nishimichi}, {Ohta}, {Shimizu}, {Takahashi}, {Takato}, {Tamura}, {Yabe}, \&
  {Yoshida}}]{Okumura:2016}
{Okumura}, T., {Hikage}, C., {Totani}, T., {et~al.} 2016, \pasj, 68, 38

\bibitem[{{Paz} {et~al.}(2008){Paz}, {Stasyszyn}, \& {Padilla}}]{Paz:2008}
{Paz}, D.~J., {Stasyszyn}, F., \& {Padilla}, N.~D. 2008, \mnras, 389, 1127

\bibitem[{{Pen} {et~al.}(2000){Pen}, {Lee}, \& {Seljak}}]{Pen:2000}
{Pen}, U.-L., {Lee}, J., \& {Seljak}, U. 2000, \apjl, 543, L107

\bibitem[{{Pereira} \& {Kuhn}(2005)}]{Pereira:2005}
{Pereira}, M.~J., \& {Kuhn}, J.~R. 2005, \apjl, 627, L21

\bibitem[{{Scannapieco} {et~al.}(2008){Scannapieco}, {Tissera}, {White}, \&
  {Springel}}]{Scannapieco:2008}
{Scannapieco}, C., {Tissera}, P.~B., {White}, S.~D.~M., \& {Springel}, V. 2008,
  \mnras, 389, 1137

\bibitem[{{Sch{\"a}fer}(2009)}]{Schafer:2009}
{Sch{\"a}fer}, B.~M. 2009, International Journal of Modern Physics D, 18, 173

\bibitem[{{Schneider} \& {Bridle}(2010)}]{Schneider:2010}
{Schneider}, M.~D., \& {Bridle}, S. 2010, \mnras, 402, 2127

\bibitem[{{Singh} {et~al.}(2015){Singh}, {Mandelbaum}, \& {More}}]{Singh:2015}
{Singh}, S., {Mandelbaum}, R., \& {More}, S. 2015, \mnras, 450, 2195

\bibitem[{{Smith} {et~al.}(2003){Smith}, {Peacock}, {Jenkins}, {White},
  {Frenk}, {Pearce}, {Thomas}, {Efstathiou}, \& {Couchman}}]{Smith:2003}
{Smith}, R.~E., {Peacock}, J.~A., {Jenkins}, A., {et~al.} 2003, \mnras, 341,
  1311

\bibitem[{{Takada} \& {White}(2004)}]{Takada:2004}
{Takada}, M., \& {White}, M. 2004, \apjl, 601, L1

\bibitem[{{Takahashi} {et~al.}(2012){Takahashi}, {Sato}, {Nishimichi},
  {Taruya}, \& {Oguri}}]{Takahashi:2012}
{Takahashi}, R., {Sato}, M., {Nishimichi}, T., {Taruya}, A., \& {Oguri}, M.
  2012, \apj, 761, 152

\bibitem[{{Tegmark} {et~al.}(1997){Tegmark}, {Silk}, {Rees}, {Blanchard},
  {Abel}, \& {Palla}}]{Tegmark:1997}
{Tegmark}, M., {Silk}, J., {Rees}, M.~J., {et~al.} 1997, \apj, 474, 1

\bibitem[{{Tenneti} {et~al.}(2015){Tenneti}, {Singh}, {Mandelbaum}, {Matteo},
  {Feng}, \& {Khandai}}]{Tenneti:2015}
{Tenneti}, A., {Singh}, S., {Mandelbaum}, R., {et~al.} 2015, \mnras, 448, 3522

\bibitem[{{The Dark Energy Survey
  Collaboration}(2005)}]{The-Dark-Energy-Survey-Collaboration:2005}
{The Dark Energy Survey Collaboration}. 2005, ArXiv Astrophysics e-prints,
  astro-ph/0510346

\bibitem[{{Tonegawa} {et~al.}(2015{\natexlab{a}}){Tonegawa}, {Totani},
  {Iwamuro}, {Akiyama}, {Dalton}, {Glazebrook}, {Ohta}, {Okada}, \&
  {Yabe}}]{Tonegawa:2015a}
{Tonegawa}, M., {Totani}, T., {Iwamuro}, F., {et~al.} 2015{\natexlab{a}},
  \pasj, 67, 31

\bibitem[{{Tonegawa} {et~al.}(2015{\natexlab{b}}){Tonegawa}, {Totani}, {Okada},
  {Akiyama}, {Dalton}, {Glazebrook}, {Iwamuro}, {Maihara}, {Ohta}, {Shimizu},
  {Takato}, {Tamura}, {Yabe}, {Bunker}, {Coupon}, {Ferreira}, {Frenk}, {Goto},
  {Hikage}, {Ishikawa}, {Matsubara}, {More}, {Okumura}, {Percival}, {Spitler},
  \& {Szapudi}}]{Tonegawa:2015}
{Tonegawa}, M., {Totani}, T., {Okada}, H., {et~al.} 2015{\natexlab{b}}, \pasj,
  67, 81

\bibitem[{{Troxel} {et~al.}(2017){Troxel}, {MacCrann}, {Zuntz}, {Eifler},
  {Krause}, {Dodelson}, {Gruen}, {Blazek}, {Friedrich}, {Samuroff}, {Prat},
  {Secco}, {Davis}, {Fert{\'e}}, {DeRose}, {Alarcon}, {Amara}, {Baxter},
  {Becker}, {Bernstein}, {Bridle}, {Cawthon}, {Chang}, {Choi}, {De Vicente},
  {Drlica-Wagner}, {Elvin-Poole}, {Frieman}, {Gatti}, {Hartley}, {Honscheid},
  {Hoyle}, {Huff}, {Huterer}, {Jain}, {Jarvis}, {Kacprzak}, {Kirk}, {Kokron},
  {Krawiec}, {Lahav}, {Liddle}, {Peacock}, {Rau}, {Refregier}, {Rollins},
  {Rozo}, {Rykoff}, {S{\'a}nchez}, {Sevilla-Noarbe}, {Sheldon}, {Stebbins},
  {Varga}, {Vielzeuf}, {Wang}, {Wechsler}, {Yanny}, {Abbott}, {Abdalla},
  {Allam}, {Annis}, {Bechtol}, {Benoit-L{\'e}vy}, {Bertin}, {Brooks},
  {Buckley-Geer}, {Burke}, {Carnero Rosell}, {Carrasco Kind}, {Carretero},
  {Castander}, {Crocce}, {Cunha}, {D'Andrea}, {da Costa}, {DePoy}, {Desai},
  {Diehl}, {Dietrich}, {Doel}, {Fernandez}, {Flaugher}, {Fosalba},
  {Garc{\'{\i}}a-Bellido}, {Gaztanaga}, {Gerdes}, {Giannantonio}, {Goldstein},
  {Gruendl}, {Gschwend}, {Gutierrez}, {James}, {Jeltema}, {Johnson}, {Johnson},
  {Kent}, {Kuehn}, {Kuhlmann}, {Kuropatkin}, {Li}, {Lima}, {Lin}, {Maia},
  {March}, {Marshall}, {Martini}, {Melchior}, {Menanteau}, {Miquel}, {Mohr},
  {Neilsen}, {Nichol}, {Nord}, {Petravick}, {Plazas}, {Romer}, {Roodman},
  {Sako}, {Sanchez}, {Scarpine}, {Schindler}, {Schubnell}, {Smith}, {Smith},
  {Soares-Santos}, {Sobreira}, {Suchyta}, {Swanson}, {Tarle}, {Thomas},
  {Tucker}, {Vikram}, {Walker}, {Weller}, \& {Zhang}}]{Troxel:2017}
{Troxel}, M.~A., {MacCrann}, N., {Zuntz}, J., {et~al.} 2017, ArXiv e-prints,
  arXiv:1708.01538

\bibitem[{{van Uitert} \& {Joachimi}(2017)}]{van-Uitert:2017}
{van Uitert}, E., \& {Joachimi}, B. 2017, \mnras, 468, 4502

\bibitem[{{Xia} {et~al.}(2017){Xia}, {Kang}, {Wang}, {Luo}, {Yang}, {Jing},
  {Wang}, \& {Mo}}]{Xia:2017}
{Xia}, Q., {Kang}, X., {Wang}, P., {et~al.} 2017, ArXiv e-prints,
  arXiv:1706.07814

\bibitem[{{Yabe} {et~al.}(2015){Yabe}, {Ohta}, {Akiyama}, {Bunker}, {Dalton},
  {Ellis}, {Glazebrook}, {Goto}, {Imanishi}, {Iwamuro}, {Okada}, {Shimizu},
  {Takato}, {Tamura}, {Tonegawa}, \& {Totani}}]{Yabe:2015}
{Yabe}, K., {Ohta}, K., {Akiyama}, M., {et~al.} 2015, \pasj, 67, 102

\bibitem[{{Zhang}(2010)}]{Zhang:2010}
{Zhang}, P. 2010, \mnras, 406, L95

\end{thebibliography}

\end{document}